\documentstyle[12pt]{article}

\def\seCtion#1{\section{#1} \setcounter{equation}{0}}
\renewcommand\theequation{\ifnum\value{section}>0{\thesection.
\arabic{equation}}\fi}
\newcommand{\be}{\begin{equation}}
\newcommand{\ee}{\end{equation}}
\newcommand{\bea}{\begin{eqnarray}}
\newcommand{\eea}{\end{eqnarray}}

\begin{document}

\pagestyle{empty}
\begin{flushright}
FIUN-GCP-07/4
\end{flushright}

\begin{center}
{\Large \bf Confining solutions of $(n+1)$-dimensional Yang-Mills
equations for flat and curved space-time with $n \le 3$}\\

\vspace{0.5cm}

{\bf J. A. S\'anchez-Monroy \footnote{jasanchezm@unal.edu.co.} and
C. J. Quimbay \footnote{cjquimbayh@unal.edu.co.,
associate researcher of CIF, Bogot\'a, Colombia.}}\\
{\it{Departamento de F\'{\i}sica}, Universidad Nacional de
Colombia\\
Ciudad Universitaria, Bogot\'a, Colombia}\\
\vspace{0.5cm} September 24, 2007
\end{center}

\vspace{0.8cm}

\begin{abstract}
We obtain exact static solutions of the $(n+1)$-dimensional $SU(3)$
Yang-Mills equations for both flat and curved space-time cases with
$n \le 3$. We find that the solutions obtained are confining
functions for $n = 1, 2, 3$. We apply the $(3+1)$ curved space-time
solution to the anti-de Sitter and Schwarzschild metrics.
\end{abstract}

\newpage

\pagestyle{plain}


\seCtion{Introduction}

The quark confinement problem \cite{conf1} can not be directly
solved by the Quantum Chromodynamics (QCD) due the perturbation
theory fails in the infrared regime \cite{conf2}. As an alternative
the color charge of the $SU(3)$ gauge fields has been related to
quark confinement \cite{ymes1}. The semiclassical approach permits
to obtain non-perturbative information about QCD starting from the
solutions of the classical partial differential equation of the
$SU(3)$ Yang-Mills theory \cite{ymes2}. To understand the role of
non-abelian gauge fields in the quark confinement problem has
motivated the search for solutions of the classical Yang-Mills field
equations in presence of static external sources. One of the first
works about this subject showed that if the external source is
distributed over a thin spherical shell the Coulomb solution is
unstable in a specific regime \cite{ymes1}. A wide range of
solutions of $(3+1)$-dimensional Yang-Mills fields equations in
presence of localized and extended external sources were obtained
later \cite{ymes2}-\cite{ymes6}. As well, some specific solutions of
the $(2+1)$-dimensional $SU(2)$ Yang-Mills equations were also
obtained \cite{d2ymes1}. On the other hand the discovery of globally
regular solutions of the Einstein-Yang-Mills equations with $SU(2)$
gauge group \cite{eymes1} originated a great interest about
spherical symmetric solutions \cite{eymes2}. Additionally, the study
of the quark confinement problem in curved space-time has been a
subject of interest \cite{barros}.

A semiclassical approach motivated in the black holes physics
technics was recently proposed to describe the energy spectra of
quarkonia by solving the Dirac equation in presence of $SU(3)$
Yang-Mills fields representing gluonic fields \cite{yu0}. In the
context of this approach, explicit calculations have shown how gluon
concentration is huge at scales of the order of $1$ fm \cite{yu1}.
The obtained solutions can model the quark confinement in a
satisfactory way suggesting that the mechanism of quark confinement
should occur within the framework of QCD \cite{yu2}. This fact
implies that the gluon fields form a boson condensate and,
therefore, gluons can be considered as classical fields \cite{yu2}.
By this reason the dynamics of the strong interaction at large
distances would be described by the equations of motion of the
$SU(3)$ Yang-Mills theory \cite{yu2}.
\par
Following this semiclassical approach, presented in detail in
\cite{yu2}, we obtain exact static solutions of the
$(n+1)$-dimensional $SU(3)$ Yang-Mills equations for both flat and
curved space-time cases with $n \le 3$. We find, in both cases,
confining solutions for $n = 1, 2, 3$. As an application of the
$(3+1)$ curved space-time solution we consider the anti-de Sitter
and Schwarzschild metrics.


\seCtion{Preliminaries}

We work in the Minkowski space $M$ in which the line element is
given by

\begin{equation}
ds^2=g_{\mu \nu}dx^{\mu}\otimes dx^{\nu},
\end{equation}
where the components $g_{\mu \nu}$ take different values depending
on the choice of coordinates and dimensions. The \emph{Hodge start
operator} $*$ is defined as: $\Lambda ^p(M)\rightarrow \Lambda
^{n-p}(M)$, where $\Lambda ^p(M)$ is the $\emph{p-form}$ over a
differentiable variety $M$ of dimension $n$. If $\{ dx^1,...,dx^n\}$
is the base for $\Lambda ^p(M)$ then
\begin{equation}
* (dx^{i_{1}}\wedge ...\wedge
dx^{i_{p}})=\frac{g^{(1/2)}}{(n-p)!}g^{i_1 l_1}...g^{i_p
l_p}\varepsilon_{l_1...l_pl_{p+1}...l_n}dx^{l_{p+1}}\wedge ...\wedge
dx^{l_{n}}.
\end{equation}
The \emph{exterior differential} $d$ is defined as: $\Lambda
^p(M)\rightarrow \Lambda ^{p+1}(M)$. This means that
\begin{equation}
d=\partial_\mu dx^{\mu}.
\end{equation}

Let $A=A_{\mu}dx^{\mu}=A_{\mu}^a \lambda _a dx^{\mu}$ be a
connection in $SU(N)$, where $\lambda _a$ are generators of the
$SU(N)$ gauge group, with $a=1, 2, ..., N^2-1$, and $A_{\mu}^a$ the
non-abelian fields. The Yang-Mills equations for $SU(N)$ can be
written using the Hodge star operator as
\begin{equation}
\label{df}
 d*F=g(*F\wedge A-A\wedge *F)+gJ,
\end{equation}
where $F=dA+A\wedge A=F^a_{\mu \nu}\lambda _a dx^{\mu}\wedge
dx^{\nu}$ is the curvature and $g$ is the coupling constant. The
non-abelian $SU(N)$ current density $J$ is given by
\begin{equation}
J=j^a_{\mu}\lambda _a*(dx^{\mu})=*j=*(j^a_{\mu}\lambda _adx^{\mu}).
\end{equation}
As an example, this current density for the case of a point particle
at rest is $J=j^a_{\mu}\lambda _a dx^{\mu}=\delta
(\vec{r})q^a\lambda _a dt$, where $q^a$ are constants and then
$q^a\lambda _a=\Upsilon$ is a constant.
\par
In a similar way as in the functional quantization of Yang-Mills
theories, we fix the gauge through the condition $div(A)=0$, or
similarly
\begin{equation}
\label{fijar} div(A)=\frac{1}{\sqrt{g}}\partial
_{\mu}(\sqrt{g}g^{\mu \nu}A_{\nu})=0.
\end{equation}

\seCtion{Solutions in a flat $(n+1)$-dimensional space-time}

The $SU(3)$ Yang-Mills equations are a non-linear system of coupled
partial differential equations. In this section we present some
exact static solutions of the $SU(3)$ Yang-Mills equations in a flat
space-time of $(n+1)$ dimensions with $n \le 3$, following the same
techniques used in \cite{yu2}. We consider the cases $n = 1, 2, 3$
and we find that in the three cases the solutions are confining
functions.
\subsection{Case $n=1$}
We first consider the Yang-Mills equations in $(1 + 1)$ dimension
using the Minkowskian metric given by
\begin{equation}
ds^2=g_{\mu \nu}dx^{\mu}\otimes dx^{\nu}=dt^2-dx^2.
\end{equation}
We assume that $A$ has the form $A = A_{\mu}(x)dx^{\mu} = A_t (x)dt+
A_x (x)dx$, where $A_{t}(x)$ is a function on $x$ (that we call
$f(x)$) that is written as a linear combination of the gauge group
generators. For this case, the gauge condition (\ref{fijar}) leads
to
\begin{equation}
\partial_x(A_x(x))= 0,
\end{equation}
being $A_x(x)$ a constant. In this coordinates, the exterior
differential is written as
\begin{equation}\label{d11}
d = \partial_t dt + \partial_xdx.
\end{equation}
For this case the Yang-Mills equations (\ref{df}) lead to
\begin{equation}
\partial^2_x(A_t(x))= \delta (x)q^a\lambda _a=\delta (x) \Upsilon,
\end{equation}
where we have taked $A_x=0$ and we have used the fact that
$*(dx)=dx$. The solution of the last equation is given by $f(x) = l
+ k|x|$, where $l$ and $k$ are constants. This solution is invariant
under parity transformation and is a linear confining solution.


\subsection{Case $n=2$}
Now we consider the Yang-Mills equations in $(2 + 1)$ dimension. We
use the Minkowskian metric in polar coordinates
\begin{equation}
ds^2=g_{\mu \nu}dx^{\mu}\otimes dx^{\nu}=dt^2-dr^2-r^2d\theta^2.
\end{equation}
If we suppose the solutions of the form $A_{\theta}(r, \theta) =
g(r)\theta + h(r)$, it is possible to see that $A_{\theta}$ does not
depend on $\theta$. In this case the gauge condition (\ref{fijar})
leads to the following equation
\begin{equation}
\partial_r(r A_r(r))= 0,
\end{equation}
being its solution given by $A_r = C/r$. If we assume that $C = 0$,
we can write $A$ in terms of $A_t = f(r)\Gamma$ and $A_r =
g(r)\Delta$, being $\Gamma$ and $\Delta$ linear combinations of the
group generators. As the exterior differential, in polar
coordinates, has the form
\begin{equation}\label{d22}
d = \partial_t dt + \partial_rdr+\partial_{\theta}d\theta,
\end{equation}
and the following relations are satisfied
 \begin{eqnarray*}
* (dt\wedge dr)&=&-rd\theta, \\
* (dt\wedge d\theta )&=&\frac{1}{r}dr,\\
* (dr\wedge d\theta )&=&\frac{1}{r}dt,
\end{eqnarray*}
then the curvature is given by
\begin{equation}\label{c22}
F=dA+gA \wedge A=-\partial _{r} f(r) \Gamma dt \wedge dr +
\partial _{r} g(r) \Delta dr \wedge d\varphi+
g f(r) g(r)[\Gamma ,\Delta ]dt \wedge d\theta,
\end{equation}
and the Hodge star operator applied over the curvature is
\begin{equation}
* F=r\partial _{r} f(r) \Gamma d\theta+\frac{\partial _{r} g(r)}{r}
\Delta dt + g \frac{f(r)g(r)}{r}[\Gamma ,\Delta ]dr.
\end{equation}
If we demand that $[\Gamma ,\Delta ]= 0$ and we consider that $*dt=r
dr \wedge d\theta$, we obtain the following differential equation
system
\begin{eqnarray}\label{222}
\partial_r(\frac{1}{r}\partial_r g(r)) \Delta &=& 0,\\
\partial _{r}(r\partial _{r} f(r))\Gamma &=&r \delta(r)\Upsilon.
\end{eqnarray}
The solutions of this equation system is given by $f(r) = d+k \log
r$ and $g(r) = d+kr^2$. The function $f(r)$ is the known confining
solution in the two dimension problem \cite{arfken}. We observe that
$g(r)$ is also a confining function. It is clear that in this case
there exists confinement and this fact is associated separately with
the temporal and spatial parts of the gluon fields. The abelian
condition $[\Gamma ,\Delta ]= 0$ is satisfied in a not trivial way
if and only if one of the following conditions holds:
\par 1) If in the combinations of $\Gamma$ and $\Delta$, in terms
of the group generators, only appears the matrix which constitute
the Cartan subalgebra of the $SU(N)$-Lie algebra. \par 2) If
$\Gamma=k \Delta$, being $k$ a constant.

\subsection{Case $n=3$}
The solutions in spherical coordinates of the $SU(3)$ Yang-Mills
equations were obtained in detail in \cite{yu0,yu1,yu2}. We only
present here the solution of $A$ given by
\begin{eqnarray}
A=A_tdt+A_{\varphi}d\varphi=(B+\frac{b}{r})\Gamma dt+(C + c r)\Delta
d \varphi,
\end{eqnarray}
which is obtained using the abelian condition $[\Gamma ,\Delta ]=
0$. We note that this solution corresponds to the Cornell potential
\cite{cornell} which has been used to describe phenomenological
features of QCD, as this is a confining function.

\seCtion{Solutions in a curved $(n+1)$-dimensional
space-time}\label{curvade}

In this section we present some static and exact solutions of the
$SU(3)$ Yang-Mills equations in a curved space-time of $(n+1)$
dimensions with $n \le 3$. We consider independently the $n = 1, 2,
3$ cases. As in the past section, the solutions for the three cases
are confining functions.

\subsection{Case $n=1$}

We consider a curved space-time in $(1 + 1)$ dimensions which metric
is given by
\begin{equation}
ds^2=g_{\mu \nu}dx^{\mu}\otimes
dx^{\nu}=\alpha^2(x)dt^2-\beta^2(x)dx^2.
\end{equation}
We assume, as in the flat $(1 + 1)$ space-time case, that $A$ has
the following functional dependence
\begin{equation}
A = A_{\mu}(x) dx^{\mu} = A_t (x) dt+ A_x(x) dx=\lambda_a f^a(x) dt
+ A_x(x) dx,
\end{equation}
where $\lambda_a$ are the group generators and the number of
functions $f^a(x)$ is the same as the number of generators. The
gauge condition (\ref{fijar}) implies that
\begin{equation}
\partial_x\left(\frac{\alpha (x) A_x(x)}{\beta(x)}\right)= 0,
\end{equation}
and then $A_x = C \beta(x)/\alpha(x)$. The exterior differential is
given by (\ref{d11}). Putting $C=0$ and using that $*(dt\wedge
dx)=\frac{1}{\alpha (x) \beta (x)} $, we obtain
\begin{equation}
*F=-\frac{\lambda_a\partial_xf^a(x)}{\alpha(x)\beta(x)}.
\end{equation}
For this case we have that $*(dt)=\frac{\beta(x)}{\alpha(x)}dx$ and
the expression (\ref{df}) leads to
\begin{eqnarray}
\partial_x\left(\frac{\lambda_a\partial_xf^a(x)}{\alpha(x)\beta(x)}
\right)dx=g\left(\frac{\partial_xf^a(x)}{\alpha(x)\beta(x)}f^b(x)
[\lambda_a,\lambda_b]\right)dt +\delta(x) \frac{\beta(x)}{\alpha(x)}
q^a\lambda_adx.
\end{eqnarray}
To have a non-trivial solution of this equation is necessary that
$[\lambda_a,\lambda_b]=0$ or alternatively
$\lambda_a\partial_xf^a(x)=f'(x)\Gamma$, where $\Gamma$ is a linear
combination of the group generators. Then we obtain the following
equation
\begin{eqnarray}
\partial_x\left(\frac{\lambda_a\partial_xf^a(x)}{\alpha(x)\beta(x)}
\right)=\delta(x) \frac{\beta(0)}{\alpha(0)}q^a\lambda_a,
\end{eqnarray}
whose solution is given by
\begin{eqnarray}
f^a(x)=k^a\int \alpha (x) \beta (x) dx +d^a,
\end{eqnarray}
where $k^a$ and $d^a$ are constants. For the case in which
$\alpha=\beta=1$ we obtain the solution of the flat space-time case,
i. e. $f(x) = d + k|x|$.


\subsection{Case $n=2$}
We consider now a curved space-time in $(2 + 1)$ dimensions given by
\begin{equation}
ds^2=g_{\mu \nu}dx^{\mu}\otimes dx^{\nu}=\alpha^2(r)dt^2-\beta^2(r)
dr^2-r^2d\theta^2.
\end{equation}
We assume that $A$ has the following functional dependence:
\begin{equation}
A = A_{\mu}(r)dxµ = A_t (r)dt+ A_r
(r)dr+A_{\theta}(r,\theta)d\theta.
\end{equation}
The gauge condition (\ref{fijar}), for this case, leads to
\begin{equation}
\partial_r\left(\frac{r \alpha ( r) A_r(r)}{\beta(r)}\right)+
\partial_{\theta}\left(\frac{\alpha ( r) \beta (r) A_{\theta}
(r,\theta)}{r^2}\right)= 0.
\end{equation}
If we discard the solutions of the form $A_{\theta}(r, \theta) =
g(r)\theta + h(r)$, we can see that $A_{\theta}$ does not depend on
$\theta$ and then the gauge condition (\ref{fijar}) leads to the
equation
\begin{equation}
\partial_r\left(\frac{r \alpha ( r) A_r(r)}{\beta(r)}\right)=0,
\end{equation}
whose solution has the form $A_r = C \frac{\beta(r)}{r \alpha(r)}$.
The exterior differential is given by (\ref{d22}). We put $C = 0$
and assume that $A$ can be written as $A_t = f(r)\Gamma$ and $A_r =
g(r)\Delta$, being $\Gamma$ and $\Delta$ linear combinations of the
group generators. If we also consider the following conditions
 \begin{eqnarray*}
 * (dt\wedge dr)&=&-\frac{r}{\alpha (r) \beta (r)} d\theta, \\
 * (dt\wedge d\theta )&=&\frac{\alpha (r) \beta (r)}{r}dr,\\
 * (dr\wedge d\theta )&=&\frac{\alpha (r)}{r \beta (r)}dt,
\end{eqnarray*}
we obtain that the curvature has the same form as the one of the
flat space-time case, which is given given by (\ref{c22}), i. e.
\begin{equation}
* F=\frac{r\partial _{r} f(r) \Gamma}{\alpha (r) \beta (r)} d\theta+
\frac{\alpha (r)}{r \beta (r)}\frac{\partial _{r} g(r)}{r} \Delta dt
+ g \frac{\alpha (r) \beta (r) f(r)g(r)}{r}[\Gamma ,\Delta ]dr.
\end{equation}
Using the expression (\ref{df}), we have that
\begin{eqnarray}  \label{eqtr}
&&\partial_r\left(r \frac{\partial_rf(r)}{\alpha(r)\beta(r)}\Gamma
\right) dr \wedge d\theta - \partial_r\left(\frac{ \alpha(r)
\partial_rg(r)}{r \beta(r)}\Delta \right) dt \wedge dr \,\, = \,\,
\delta (\vec{r}) \frac{r \beta(r)}{\alpha(r)} \Upsilon dr
\wedge d\theta \nonumber \\
&+&g^2\left( \frac{f(r)g(r)^2\alpha(r)\beta(r)}{r}[[\Gamma,\Delta],
\Delta] dr \wedge d\theta - \frac{f(r)^2g(r)\alpha(r)\beta(r)}{r}
[[\Gamma,\Delta],\Gamma] dt \wedge dr
 \right).\nonumber\\
\end{eqnarray}
If we demand that $[\Gamma ,\Delta ]= 0$, equation (\ref{eqtr})
leads to the following equation system
\begin{eqnarray}
\partial_r \left(\frac{\alpha(r)}{r \beta (r)}\partial_r g(r)
\right) &=& 0,\\
\partial _{r}\left( \frac{r}{\alpha (r) \beta (r)}\partial _{r}
f(r) \right)&=& \delta (\vec{r}) \frac{r \beta(0)}{\alpha(0)}
\Upsilon.
\end{eqnarray}
The solution of this equation system is given by
\begin{eqnarray}
g(r)&=&k_2 \int \frac{r \beta (r)}{\alpha (r)}dr+d_2,\\
f(r)&=&k_1 \int \frac{\alpha (r) \beta (r)}{r}dr+d_1.
\end{eqnarray}
For the case $\alpha=\beta=1$ we obtain the solution of the $(2 +
1)$ flat space-time case, i. e. $g(r) = d_2+k_2r^2$ and $f(r) =
d_1+k_1 \log r$.


\subsection{Case $n=3$}
The metric for a curved static space-time and spherically symmetric
can be specified by
\begin{equation}
ds^2=g_{\mu \nu}dx^{\mu}\otimes dx^{\nu}=\alpha^2(r)dt^2-
\beta^2(r)dr^2-r^2(d \theta ^2+sin^2\theta d\varphi ^2).
\end{equation}
We assume that $A$ has the following functional dependence
\begin{equation}
A=A_{\mu}(r)dx^{\mu}=A_{t}(r)dt+A_{r}(r)dr+A_{\theta}(r)d\theta+
A_{\varphi}(r)d\varphi.
\end{equation}
For this case, the gauge condition (\ref{fijar}) leads to
\begin{equation}
\partial_r\left(\frac{r^2 \alpha ( r) A_r(r)}{\beta(r)}\right)+
\alpha ( r) \beta (r) A_{\theta}(r)cot\theta= 0,
\end{equation}
and then, $A_{\theta}=0$ and $A_r = C \frac{\beta(r)}{r^2
\alpha(r)}$. We can put $C=0$ in the $A_{r}$ solution because this
does not affect the form of $A_{t}$ and $A_{\varphi}$ solutions. Now
we take $A_{t}=f(r)\Gamma$ and $A_{\varphi}=g(r)\Delta$, being
$\Delta$ and $\Gamma$ linear combination of the group generators. As
the exterior differential in spherical coordinates is given by
\begin{equation}
d=\partial_tdt+\partial_rdr+\partial_{\theta}d\theta+
\partial_{\varphi}d\varphi,
\end{equation}
then the curvature is given by
\begin{equation}
F=dA+gA \wedge A=-\partial _{r} f(r) \Gamma dt \wedge dr +
\partial _{r} g(r) \Delta dr \wedge d\varphi+g f(r) g(r)
[\Gamma ,\Delta ]dt \wedge d\varphi.
\end{equation}
Applying the Hodge start operator (\ref{df}) over $F$ and using the
following relations
 \begin{eqnarray*}
 * (dt\wedge dr)&=&-\frac{r^2  sin \theta}{\alpha(r)^2 \beta(r)^2}
 d\theta \wedge d \varphi, \\
 * (dt\wedge d\theta )&=&\frac{sin \theta}{\alpha(r)^2} dr \wedge d \varphi,\\
 * (dt\wedge d\varphi)&=&\frac{-1}{\alpha (r)^2 sin\theta }dr \wedge d \theta, \\
 * (dr\wedge d\varphi )&=&\frac{-1}{\beta (r)^2 sin\theta } dt \wedge d\theta ,\\
 * (dr\wedge d\theta )&=& \frac{sin \theta}{\beta(r)^2} dt \wedge d \varphi ,\\
 * (d\theta \wedge d\varphi )&=&\frac{1}{r^2sin\theta } dt \wedge dr,
\end{eqnarray*}
we obtain, for $r\not=0$, the following equation system

\begin{eqnarray}\label{23}
\partial _{r}  \left( \frac{\partial_r g(r)}{\beta (r)^2 } \right) \Delta &=&
\frac{g^2}{\alpha(r)^2} f(r)^2 g(r)[\Gamma ,[\Gamma ,\Delta ]],\\
\partial _{r}\left(\frac{r^2\partial _{r} f(r)}{\alpha(r)^2 \beta(r)^2} \sin ^2
\theta\right) \Gamma &=&\frac{g^2 f(r)g(r)^2}{\alpha(r)^2}[\Delta
,[\Gamma ,\Delta ]].
\end{eqnarray}
As for the flat space-time case, we demand this equation system to
satisfy the condition $[\Delta ,[\Gamma ,\Delta ]]=0$ and then we
obtain
\begin{eqnarray}
\partial _{r}  \left(\frac{\partial_r g(r)}{\beta (r)^2 } \right)&=&0,\\
\partial _{r}\left(\frac{r^2\partial _{r} f(r)}{\alpha(r)^2 \beta(r)^2}\right)&=&0.
\end{eqnarray}
The solutions of these two equation are
\begin{eqnarray}
g(r)&=&b_1 \int \beta(r)^2dr+B_1, \label{25}\\
f(r)&=&a_1 \int \frac{\alpha(r)^2 \beta(r)^2}{r^2}dr+A_1. \label{24}
\end{eqnarray}
We observe that for $\alpha=\beta=1$, the solutions (\ref{25}) and
(\ref{24}) allow to the confining solutions of the flat space-time
case given by +$g(r)=b_1 r +B_1$ and $f(r)= -a_1/r +A_1$.\par

As a particular application of the $(3 + 1)$ case solution we
consider now the anti-de Sitter metric given by
\begin{equation}
ds^2=(1-\Lambda r^2/3)dt^2-(1-\Lambda r^2/3)^{-1}dr^2-r^2 (d \theta
^2+\sin^2\theta d\varphi ^2).
\end{equation}
Because $\alpha(r)$ is the inverse of $\beta(r)$, then the Coulomb
solution has not deformations respect to the flat space-time case
and it is given by $f(r)=a_1/r+A_1$. The linear solution $g(r)$
changes respect to the flat case. We obtain the solution explicitly
in the the two following situations $\Lambda>0 $ and $\Lambda <0 $,
so
\begin{equation}
g(x) = \left\{
\begin{array}{cl}
b_1 \tanh ^{-1}\left(\frac{r \Lambda^{1/2}}{3^{1/2}}\right)
+B_1,&\mbox{if } \Lambda >0,\\
b_1 \tan ^{-1}\left(\frac{r (-\Lambda)^{1/2}}{3^{1/2}}\right)
+B_1,&\mbox{if } \Lambda <0,\\
\end{array}\right.
\end{equation}

The function $g(r)$, for the limit cases $|\Lambda|<<1$ and $r<<1$,
has the form $g(r)\simeq b_1r+B_1$, recovering the the flat
space-time case behavior.
\par
Another application for the $(3 + 1)$ curved space-time solution is
the Schwarzschild metric given by
\begin{equation}\label{shr}
ds^2=(1-2M/r)dt^2-(1-2M/r)^{-1}dr^2-r^2(d \theta ^2+ sin^2\theta
d\varphi ^2).
\end{equation}
As happened in the first application, the Coulomb solution has not
deformations respect to the flat space-time case, but the linear
solution has. The function $g(r)$ for this case is
\begin{equation}
g(r)=b_1 (r + 2 M \ln|r-2 M|)+B_1
\end{equation}
This solution, for the limit $r>>M$, has the form $g(r)\simeq
b_1r+B_1$, recovering the flat space-time case behavior.


\seCtion{Summary}

\hspace{3.0mm}

We have presented some exact static solutions for the $SU(3)$
Yang-Mills equations in a flat and a curved space-time of $(n+1)$
dimensions with $n \le 3$. We have found in both cases there are
confining solutions for $n = 1, 2, 3$. For the $(1+1)$ case, both in
the flat and curved space-time, we found that the solution for the
temporary part can be written as $A_t=f(r)\Gamma$. To find analytic
solutions in the $(2+1)$ case is necessary to demand the abelian
condition given by $[\Delta,\Gamma]=0$. For the cases $(1+1)$ and
$(3+1)$ this condition is satisfied naturally. We presented in
detail the solution for $(3+1)$ curved space-time case and we
applied this solution to the anti-de Sitter and Schwarzschild cases.
In both cases the Coulomb solution does not have deformations
respect to the flat space-time case, while in the linear solution
there exists deformation. As a perspective it would be interesting
to understand the role of the confining solutions in a model of
relativistic quark confinement in low dimensionality.

\hspace{3.0mm}

\section*{Acknowledgments} We are indebted to Maurizio de Sanctis
and Rafael Hurtado for important suggestions about the elaboration
of the present work. C. Quimbay would like to thank High Energy
Latinamerican European Network (HELEN) for financial support.


\begin{thebibliography}{999}

\bibitem{conf1} A. Chodos, R. L. Jaffe, K. Johnson, C. B. Thorn and V. F. Weisskopf,
Phys. Rev. D9 (1974) 3471; Phys. Rev. D10 (1974) 2599; K. G. Wilson,
Phys. Rev. D10 (1974) 2445; Y. Nambu, Phys. Rev. D10 (1974) 4262; C.
G. Callan, R. Dashen and D. J. Gross, Phys. Rev. D17 (1978) 2717.
\bibitem{conf2} W. J. Marciano and H. Pagels, Phys. Rept. 36 (1978)
137.
\bibitem{ymes1} J. E. Mandula, Phys. Rev. D14 (1976) 3497.
\bibitem{ymes2} R. Jackiw, L. Jacobs and C. Rebbi, Phys. Rev. D20 (1979) 474; R. Jackiw
and P. Rossi, Phys. Rev. D21 (1980) 426.
\bibitem{ymes3} P. Sikivie and N. Weiss, Phys. Rev. Lett. 40 (1978) 1411; Phys.
Rev. D18 (1978) 3497; Phys. Rev. D20 (1979) 487.
\bibitem{ymes4} D. Horvat and K. S. Viswanathan, Phys. Rev. D23 (1981)
937; D. Horvat, Phys. Rev. D34 (1986) 1197.
\bibitem{ymes5} R. Teh, W. K. Koo and C. H. Oh, Phys. Rev. D23 (1981)
3046; C. H. Oh, R. Teh and W. K. Koo, Phys. Rev. D24 (1981) 2305;
Phys. Rev. D25 (1982) 3263; C. H. Oh, Phys. Rev. D25 (1982) 2194; J.
Math. Phys. 25 (1984) 660; Phys. Rev. D47 (1993) 1652; C. H. Oh, S.
N. Show and C. H. Lai, Phys. Rev. D30 (1984) 1334; C. H. Oh, C. H.
Lai and C. P. Soo, Phys. Rev. D32 (1985) 2843.
\bibitem{ymes6} K. Cahill, Phys. Rev. Lett. 41 (1978) 599; U. Sarkar and
Raychaudhuri, Phys. Rev. D26 (1982) 2804; H. Arodz, Nucl. Phys. B207
(1982) 288; Acta Phys. Polon. B14 (1983) 825; Phys. Rev. D35 (1987)
4024; F. Nill, Ann. Phys. (N.Y.) 149 (1983) 303; G. K. Savvidy,
Phys. Lett. B130 (1983) 303; S. J. Chang, Phys. Rev. D29 (1984) 259;
S. K. Paul and A. Khare, Phys. Lett. B138 (1984) 402; D. Sivers,
Phys. Rev. D34 (1986) 1141; E. Malec, Acta Phys. Pol. B18 (1987)
1017.
\bibitem{d2ymes1} E. D'Hoker and L. Vinet, Ann. Phys. (N.Y.) 162 (1985) 413;
C. H. Oh, L. H. Sia and R. Teh, Phys. Rev. D40 (1989) 601.
\bibitem{eymes1} R. Bartnik and J. McKinnon, Phys. Rev. Lett. 61 (1988) 141.
\bibitem{eymes2} P. Bizon, Phys. Rev. Lett. 64 (1990) 2844; M. S. Volkov
and D. V. Gatsov, Phys. Rept. 319 (1999)1; E. Winstanley, Class.
Quant. Grav. 16 (1999) 1963; Z. Zou, J. Math. Phys. 42 (2001) 1085;
A. G. Wasserman, J. Math. Phys. 41 (2000) 6930; Y. Brihaye, A.
Chakrabarti and D. H. Tchrakian, Class. Quant. Grav. 20 (2003) 2765;
Y. Brihave and B. Hartmann, Class. Quant. Grav. 22 (2005) 183; Y.
Brihaye, E. Redu and D. H. Tchrahian, Phys. Rev. D75 (2007) 024022.
\bibitem{barros} C. C. Barros Jr., Eur. Phys. J. C45 (2006) 421.
\bibitem{yu0} Y. P. Goncharov, Mod. Phys. Lett. A16 (2001) 557;
Europhys. Lett. 62 (2003) 684; Y. P. Goncharov and E. A. Choban,
Mod. Phys. Lett. A18 (2003) 1661; arXiv:hep-th/0512099 v1 8 Dec
2005; Phys. Lett. B641 (2006) 237.
\bibitem{yu1} Y. P. Goncharov and A. A. Bytsenko, Phys. Lett. B602 (2004) 86.
\bibitem{yu2}  Y. P. Goncharov, Phys. Lett. B 617 (2005) 67.
\bibitem{arfken} G. B. Arfken and H. J. Weber, \emph{Mathematical methods for
physicists}, fifth edition, Academic Press (2001).
\bibitem{cornell} E. Eichten, K. Gottfried, T. Kinoshita, K. D. Lane, and T. M.
Yan, Phys. Rev. D17 (1978) 3090.

\end{thebibliography}
\end{document}